# Anisotropy of Magnetization Reversal and Magnetoresistance in Square Arrays of Permalloy Nano-Rings

A.V.Goncharov, A.A.Zhukov, V.V.Metlushko, G.Bordignon, H.Fangohr, C.H.de Groot, J.Unguris,
W.C.Uhlig, G.Karapetrov, B.Ilic, P.A.J.de Groot

*Abstract*—**Magnetization reversal mechanisms and impact of magnetization direction are studied in square arrays of interconnected circular permalloy nanorings using MOKE, local imaging, numerical simulations and transport techniques.**

## I. INTRODUCTION

Magnetic nano-structures demonstrate a wealth of magnetic properties, which are extremely sensitive to the shape and the order of nano-elements strongly interacting via exchange and dipolar coupling [1]. From a technological point of view, such materials are important for applications in magnetic recording and, in particular, for rapidly developing magnetic random access memory (MRAM) devices [2]. Magnetic nano-rings are among the most promising candidates for the future MRAM applications [3]. Their main advantage stems from the presence of a hole in the center of the ring, which eliminates the highly energetic vortex core existing in a dot element. Magnetic properties of separate nano-rings have been reported in several publications (e.g., [4-6]). In this work we study square arrays of connected circular permalloy rings to explore magnetization reversal and magnetoresistance mechanisms in dependence on a magnetic field direction.

## II. EXPERIMENTAL PROCEDURE

Periodic square arrays of 25nm thick permalloy nano-rings (Fig.1) have been prepared by e-beam lithography on a (100) silicon wafer spin-coated with bi-layer resists (PMMA and P(MMA-MAA) copolymer). Using e-beam evaporation in a high vacuum a permalloy film was deposited, and after a lift-off in acetone the arrays of magnetic permalloy rings were obtained. In this work we study square arrays with a period of 1000nm of connected circular rings having external and internal diameters of 1100nm and 650nm, respectively. Magneto-optical Kerr effect (MOKE) magnetometry, magnetoresistance, scanning electron microscopy with polarization analysis (SEMPA) [7] and magnetic force microscopy (MFM) were used for characterization. Measurements of *M-H* curves were performed at room temperature for magnetic fields in the film plane with various angles between the field direction and an edge of the square array. The transport measurements were made by a four probe method for different in-plane and out-off-plane magnetic field directions. Magnetic simulations have been made using OOMMF software suite [8].

**FIG. 1 HERE**

## III. RESULTS AND DISCUSSION

### A. In-plane magnetization reversal anisotropy

For magnetic field directions close to the main axis of the square array we observe sharp one step magnetization switching (Fig.2a). More complicated behavior is observed near the 45° type directions (Fig2b). As can be seen from Fig.3 the angular dependence of in-plane coercive fields demonstrates significant variations. Close to the main directions of square arrays ($\varphi$=0° and 90°) the coercive field roughly follows a fixed projection of magnetic field on the corresponding main direction. Near the 45° type directions this behavior changes and a pronounced minimum is reached.

**FIG. 2 HERE**

Manuscript received March 13, 2006. This work was supported in part by the U.K. Engineering and Physical Sciences Research Council and the USA National Science foundation.

A.V.Goncharov, A. A. Zhukov and P. A. J. de Groot are with the School of Physics and Astronomy, University of Southampton, SO17 1BJ UK (e-mail: avg@phys.soton.ac.uk; aaz@phys.soton.ac.uk; pajdeg@phys.soton.ac.uk)

V.V.Metlushko is with the Department of Electrical and Computer Engineering, University of Illinois at Chicago, Chicago, IL 60607-0024, USA (e-mail: vmetlush@ece.uic.edu)

G.Bordignon and H. Fangohr are with the School of Engineering Sciences, University of Southampton, SO17 1BJ UK (e-mail: gb2@soton.ac.uk; H.FANGOHR@soton.ac.uk)

C.H.de Groot is with the School of Electronics and Computer Science, University of Southampton,SO17 1BJ UK (e-mail: chdg@ecs.soton.ac.uk)

J.Unguris and C.Uhlig are with National Institute of Standards and Technology, 100 Bureau Drive Stop 8412, Gaithersburg, MD 20899-8412, USA (e-mail: unguris@nist.gov; uhlig@nist.gov)

G. Karapetrov is with the Materials Science Division, Argonne National Laboratory, 9700 South Cass Ave., Argonne, IL60439, USA (e-mail: goran@anl.gov)

B.Ilic is with the Cornell Nanofabrication Facility and School of Applied and Engineering Physics, Cornell University, Ithaca, NY 14853 (rob@cnf.cornell.edu)

## FIG. 3 HERE

The origin of the angular anisotropy observed is clearly related to the connections between rings, which break the circular symmetry. For the soft permalloy the material anisotropy is a minor factor and the coercive field should be determined by the interplay of dipolar and exchange interactions. The latter factor dominates in connected circular rings resulting in a larger coercivity for magnetic fields along the connections. This differs from square rings where the shape related dipolar term becomes more important and a larger coercive field is reached for the 45 degree direction [9].

## FIG. 4 HERE

Our numerical simulations qualitatively reproduce the experimental behavior. The remanent state according to this analysis is an onion state (Fig.4a) and the reversal mechanisms are realised by $90^o$ and $180^o$ rotations. Our SEMPA and MFM imaging confirms the presence of the onion state in connected square ring arrays (Fig.4b, c). A clear rhombic like domain can be seen at interconnections between the onion states. Such magnetic structures appear similar to observations in chains of circular Co rings [10].

### B.  Magnetoresistance for different in-plane and out-of-plane magnetic field directions

For magnetic fields parallel to the film plane the magnetoresistance ($MR = R(H) / R(0) - 1$) demonstrate an AMR peak at small fields. For magnetic fields tilted out-of-plane two sharp peaks are superimposed on a characteristic $\sim H^2$ AMR behaviour. As can be seen from figure 5 the sign of these peaks is dependent on the direction of the in-plane magnetic field component $H_{//}$. It is negative for $H_{//}$ parallel to the electric current and positive for the transverse orientation.

## FIG. 5 HERE

Using a rotational stage with high angular resolution, the magnetoresistive curves have been measured for various directions of the magnetic field with respect to the sample plane. The electric current has been applied along an edge of the square array. As can be seen from Fig.6 we find that for angles smaller than 80° from the film plane the position of peaks corresponds to an in-plane magnetic field of ~17 mT close to the value of in-plane coercive field. On the basis of numerical simulations and MFM imaging we relate the peaks to 90 degree switching of the onion state.

## FIG. 6 HERE

## IV.  CONCLUSIONS

Magnetization reversal mechanisms and the impact of magnetization direction are studied in square arrays of connected circular permalloy nanorings using MOKE, SEMPA, MFM, numerical simulations and transport techniques. Angular dependences of in-plane coercive fields demonstrate significant variations reflecting impact of inter-ring connections. Magnetoresistance acquires two pronounced peaks with a sign dependent on the current direction and position defined by an in-plane component of the field.

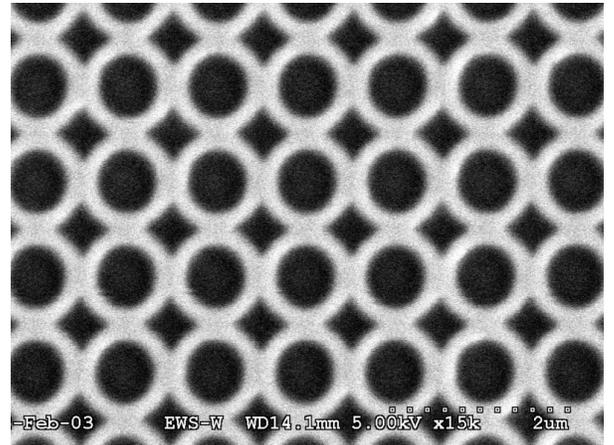

Fig. 1. SEM image of a permalloy array with interconnected circular nano-rings.

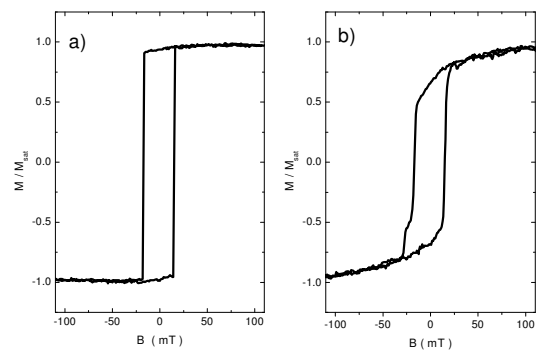

Fig. 2. MOKE magnetization loops for φ = 0⁰ (a) and φ = 45° (b). The φ = 0 direction corresponds to the magnetic field along an edge of a square lattice.

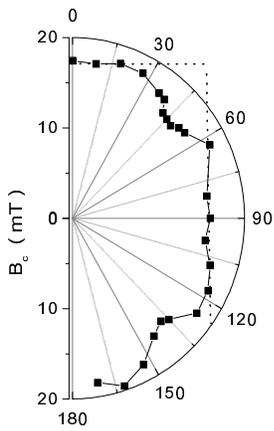

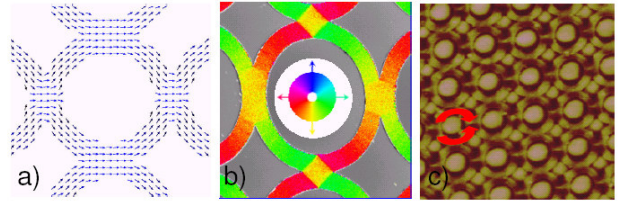

Fig.4. Moment distribution from numerical simulations (a), SEMPA (b) and MFM (c) for the remanent state after field applied along the x-axis.

Fig. 3. Angular dependence of coercivity of this array on the direction of the in-plane applied magnetic field. Dashed lines correspond to a fixed projection onto the main directions.

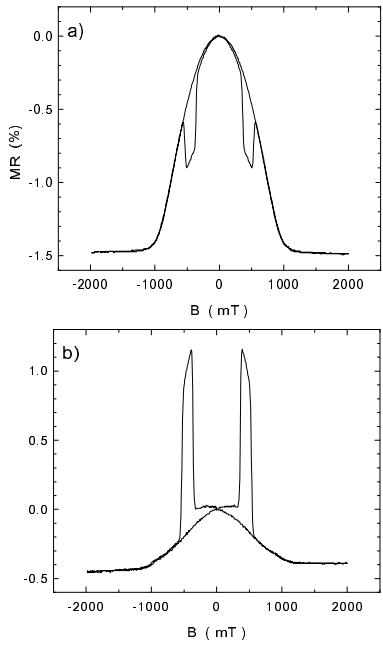

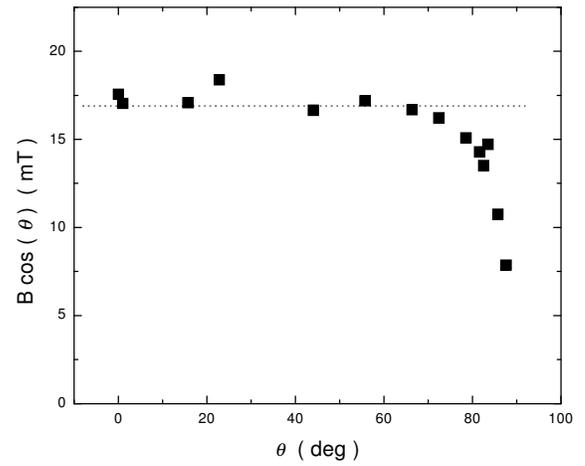

Fig.6. Angular dependence of in-plane component of peak fields. Zero angle θ corresponds for the in-plane magnetic field.

Fig. 5. Magnetoresistance of the interconnected circular ring array (Fig.1) for $H_{//}$ component parallel (a) and transverse (b) to the electric current (directed along the edge of the square array).